\documentclass{iaus}
\usepackage{graphicx}
\usepackage{epsf}
\usepackage{psfig}
\usepackage{graphics}
\usepackage{lscape}

\title[Accurate AGN black hole masses and the scatter in the $M_\bullet$ -- $L_{bulge}$ relationship] 
{Accurate AGN black hole masses and the scatter in the $M_\bullet$ -- $L_{bulge}$ relationship}

\author[C. Martin Gaskell]   
{C. Martin Gaskell}

\affiliation{Astronomy Department, University of Texas, Austin, TX 78712-0259 \\email: {\tt gaskell@astro.as.utexas.edu}}

\pubyear{2010}
\volume{267}  
\pagerange{1--1}
\jname{Co-Evolution of Central Black Holes and Galaxies}
\editors{B.M.\ Peterson, R.S.\ Somerville, \& T.\ Storchi-Bergmann, eds.}
\begin{document}

\maketitle

\begin{abstract}
A new empirical formulae is given for estimating the masses of black holes in AGNs from the H$\beta$ velocity dispersion and the continuum luminosity at 5100~\AA\.~  It is calibrated to reverberation-mapping and stellar-dynamical estimates of black hole masses.  The resulting mass estimates are as accurate as reverberation-mapping and stellar-dynamical estimates.  The new mass estimates show that there is very little scatter in the $M_\bullet$ -- $L_{bulge}$ relationship for high-luminosity galaxies, and that the scatter increases substantially in lower-mass galaxies.


\end{abstract}

Accurate AGN black hole masses, $M_\bullet$, can be estimated from the velocity dispersion of the broad H$\beta$ line, $\sigma_{H\beta}$, and the luminosity at 5100~\AA, $L_{5100}$, by the equation:

\begin{equation}
\log M_\bullet = 1.65 \log(\sigma_{H\beta}/1000) +  0.615 (\log{L_{5100}}-44) + 7.63 ,
\end{equation}

These masses agree with reverberation-mapping masses to $\pm 0.22$ dex.  This suggests that the masses are determined by the new empirical relationship to $\pm 0.16$ dex.  Fig.\@ 1 shows the dispersion about the $M_{bh}$--$L_{bulge}$ relationship as a function of $L_{bulge}$ for 34 AGNs (in equal bins). Note the very small scatter for the most luminous galaxies.  \cite{Gaskell09} shows that the dispersion in the relationship between $M_\bullet$ and stellar velocity dispersion also increases with decreasing bulge luminosity.

\vspace{-0.5cm}
\begin{figure}[b!]
\begin{center}
\epsfxsize = 65 mm
\epsfbox{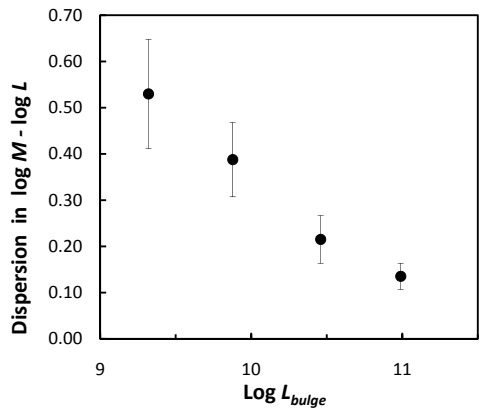}
\caption{The scatter in the AGN $M_\bullet$ -- $L_{bulge}$ relationship as a function of bulge luminosity
\label{fig1}
}
\end{center}
\end{figure}

\end{document}